\RequirePackage{lineno} 
\documentclass[12pt]{elsarticle}

\usepackage{graphicx}
\usepackage{epsfig}
\usepackage{optional}
\usepackage{amsmath}
\usepackage{amssymb}

\begin{document}

\linenumbers


\title{A Bayesian approach to evaluate confidence intervals in counting 
experiments with background}

\author{F.~Loparco}
\address{Dipartimento Interateneo di Fisica ``M.~Merlin'',
dell'Universit\`a degli Studi di Bari and INFN Sezione di Bari,
Via Amendola 173, I-70126 Bari, Italy}
\ead{loparco@ba.infn.it}
\author{M.~N.~Mazziotta}
\address{INFN Sezione di Bari,
Via Amendola 173, I-70126 Bari, Italy}
\ead{mazziotta@ba.infn.it}

\date{\today}

\begin{abstract}

In this paper we propose a procedure to evaluate Bayesian confidence 
intervals in counting experiments where both signal and background 
fluctuations are described by the Poisson statistics. The results
obtained when the method is applied to the calculation of upper 
limits will also be illustrated. 

\end{abstract}

\begin{keyword}
Counting experiments \sep Confidence intervals
\sep Upper limits 
\end{keyword}

\maketitle

\section{Introduction}
\label{sec:intro}

The evaluation of confidence intervals and limits is a common 
task in particle physics~\cite{pdg}. Usually the goal of an 
experiment is that of determining a parameter $\theta$ starting
from a set of measurements of a random variable $x$ (the outcome
of the experiment) and assuming an hypothesis for the probability 
distribution function (p.d.f.) $f(x | \theta)$. 
A confidence interval for $\theta$ at the confidence level 
(or coverage probability) $1-\beta$ includes the true value 
of $\theta$ with a probability $1 - \beta$. This means that
if the experiment were repeated many times, the estimates of
$\theta$ will fall in the confidence interval in a fraction
$1 - \beta$ of the experiments
\footnote{In the most general case $\theta$ 
is a vector of parameters, i.e. 
$\theta = (\theta_{1},\theta_{2}, \ldots, \theta_{n})$
and $x$ represents a set of observables, i.e.
$x = (x_{1}, x_{2}, \ldots, x_{m})$. Consequently, the p.d.f.
$f(x | \theta) = f(x_{1}, \ldots, x_{m} | \theta_{1}, \ldots, \theta_{n})$ 
will be a function of $m+n$ variables. Also, the confidence
interval defined in the case of a single parameter 
becomes a $n$-dimensional
confidence region for the vector of parameters $\theta$.}.

Confidence intervals can be evaluated following 
either the frequentist~\cite{neyman,feldman,rolke1} 
or the Bayesian approaches (a detailed discussion 
about these methods can be found in ref.~\cite{pdg}).
One of the main issues arising when applying
the frequentist approach is that the confidence
intervals may include unphysical regions for the
parameter. These cases can be fixed by introducing
some ``ad-hoc'' corrections in the mathematical 
procedures used to evaluate the intervals
(see for instance ref~\cite{rolke1,rolke2}). 
Such issues can be easily avoided when following 
the Bayesian approach, by properly choosing the
prior p.d.f. 
\footnote{This can be done by assigning null 
probabilities to the values of $\theta$ in the
unphysical regions.}
for the parameter.
However, the fact that the Bayesian approach 
requires a choice of the prior p.d.f. for the
parameter introduces some degree of arbitrariness
in the evaluation of confidence intervals. 

In many cases the outcome of an experiment can be
described in terms of a set parameters, not all
being of any interest for the final result 
(nuisance parameters). In these cases 
the experimenter wishes to evaluate
confidence regions on the parameters of interest,
in a manner that is independent 
on the nuisance parameters.
From a mathematical point of view, the Bayesian 
approach allows one to treat the nuisance
parameters in a very simple and straightforward way. 
Indicating respectively with $\theta$ and $\nu$ the 
parameters of interest and the nuisance parameters,
one has to write down their joint prior 
p.d.f. $\pi(\theta, \nu)$
\footnote{Hereafter we will us the
greek symbol $\pi$ to indicate prior p.d.f.s and the 
latin symbol $p$ to indicate posterior p.d.f.s}
and to evaluate from it the marginal 
prior p.d.f. for $\theta$. 
In most cases $\theta$ and $\nu$ 
are independent random variables, and
their joint p.d.f. can be factorized as 
$\pi(\theta, \nu) = \pi(\theta) \pi(\nu)$,
thus making the calculation easier. 

In the following sections we will illustrate an
application of the Bayesian approach to the analysis 
of a counting experiment with background. We will
assume that the outcome of the experiment can be 
modeled in terms of a parameter of interest (signal) 
and of a nuisance parameter (background), that will
be supposed to be independent on each other. The
posterior p.d.f. for the signal will be derived
starting from a set of simple and quite 
general assumptions on the prior p.d.f.s for
both the signal and the background. The formulas 
to evaluate upper limits on the signal will also
be derived and the results will be discussed. 
Our results are a generalization of the ones
illustrated in ref.~\cite{pdg}, where the same
problem is discussed, but without taking background
fluctuations into account.

\section{Formulation of the problem}
\label{sec:problem}

Let us consider a counting experiment in which one wants to measure
the signal counts in presence of a background, that is measured 
in a subsidiary experiment. 
Such a situation can happen when measuring the activity of a 
weak radioactive source in presence of background with a counting 
device, like a Geiger-Muller detector. In this case two independent 
measurements are carried out, one for the signal and one for the 
background, and the results have to be properly combined.
Another example is an experiment in which the signal and the 
background are evaluated looking at two different space regions.
Also in this case, the signal is evaluated by combining
the counts measured in the signal and background regions
\footnote{We prefer to use the word ``combine'' instead 
of ``subtract'' because in many cases the signal cannot be 
evaluated by simply subtracting the counts in the background 
region from the ones in the signal region. This may happen for 
instance when the counts in the background region are more than 
the ones in the signal region.}.

In the following we shall denote with $n$ and $m$ the number of
counts measured respectively in the signal and in the background
regions. We will also indicate with $s$ and $b$ the true 
values of the signal and of the background counts respectively,
and we will assume that the true value of the background counts
in the signal region is given by $cb$, where $c$ is a 
constant value that is assumed to be exactly known. 
Let us consider, as a first example, an experiment to measure 
the activity of a radioactive source. 
In this case $n$ and $m$ are respectively the counts 
recorded during a time interval $T_{s}$ in presence of the source, 
and during a time interval $T_{b}$ when the source has been removed. 
The value of $c$ can be estimated as $c=T_{s}/T_{b}$, and the source 
activity can be evaluated as $s/T_{s}$ once the value of $s$
has been measured. A second example, taken from the astrophysics, 
is the measurement of the photon flux of a point source in 
the sky with a gamma-ray detector. In this case 
the signal region can be a cone centered on the source 
direction with a given angular aperture, while the background 
region can be an annulus far away from the source. 
In this case $c$ can be evaluated as the ratio between 
the solid angles of the signal region and of the background 
region, eventually multiplied for the ratio between the live 
times of the two regions.

Under the above assumptions, the probability of
measuring $m$ counts in the background region will be a Poisson
distribution with mean value $b$, i.e.:

\begin{equation}
\label{eq:bkgpdf}
p(m|b) = \textrm{e}^{-b}~\cfrac{b^{m}}{m!}
\end{equation}
while the probability of measuring $n$ counts in the signal region
will be a Poisson distribution with mean value $s+cb$, i.e.:

\begin{equation}
\label{eq:signalpdf}
p(n|s,b) = \textrm{e}^{-(s+cb)}~\cfrac{(s+cb)^{n}}{n!}
\end{equation}
Since the two measurements are independent, the joint p.d.f.
for $n$ and $m$ will be given by:

\begin{equation}
\label{eq:jointpdf}
p(n,m|s,b) = 
\textrm{e}^{-(s+cb)}~\cfrac{(s+cb)^{n}}{n!} 
~\textrm{e}^{-b}~\cfrac{b^{m}}{m!}
\end{equation}

Our problem is that of evaluating a Bayesian confidence interval 
(or an upper limit) for the parameter $s$, independently 
on $b$. An analogous problem is discussed in the textbook~\cite{cowan}
and in ref.~\cite{pdg}, where the background value is assumed to be 
exactly known. Our discussion will be therefore a generalization 
of refs.~\cite{pdg,cowan}. A possible solution of the problem taking
background fluctuations into account is given in ref.~\cite{helene}.
However, in ref.~\cite{helene}, a gaussian p.d.f. is
used to model the background, the results being
valid in the case of large counts. A similar problem is also illustrated 
in ref.~\cite{heinrich}, where Bayesian confidence interval are evaluated 
for a Poisson signal with known background and with fluctuations 
on the detection efficiency. 

Another possible solution to our problem is also given in 
refs.~\cite{rolke1,rolke2}, where a frequentist approach is
followed with the application of the profile 
likelihood method. However, while the procedure described in 
refs.~\cite{rolke1,rolke2} requires some adjustments 
to handle the cases when $n<cm$ or when either $n=0$ or $m=0$, 
in our method the treatement of these cases is straightforward
and does not require any adjustment. In fact, 
the formulas that will be derived in sections~\ref{sec:bayes}
and~\ref{sec:upperlimits} are valid for all the values of $n$
and $m$.

\section{The Bayesian approach}
\label{sec:bayes}

The implementation of the Bayesian approach requires 
the ``probabilistic inversion'' of eq.~\ref{eq:jointpdf},
i.e. the evaluation of the conditional p.d.f. $p(s,b|n,m)$
starting from $p(n,m|s,b)$ and applying the Bayes'theorem.

\subsection{Choice of the priors}
\label{sec:priors}

The application of the Bayes'theorem starts from the assumption 
of a prior p.d.f. for both the random variables $s$ and $b$. 
In the following we will assume that $s$ and $b$ are independent.

For the true background value $b$ we will assume a 
uniform prior:

\begin{equation}
\label{eq:bkgprior}
\pi(b) = \left\{ 
\begin{array}{ll}
\pi_{0} & b\geq 0\\ 
0 & b<0
\end{array}
\right.
\end{equation} 
where $\pi_{0}>0$ is a constant. The parameter $b$ 
is only constrained to be non-negative. 

On the other hand, for the signal true value $s$ we
will assume a prior p.d.f. given by:

\begin{equation}
\label{eq:signalprior}
\pi(s) = \left\{ 
\begin{array}{ll}
ks^{-\alpha} & s\geq 0\\ 
0 & s<0
\end{array}
\right.
\end{equation} 
with $k>0$. Also in this case, the only constraint 
on the parameter $s$ is that it must be non-negative.
 
It is worth to point out at this stage that both 
$\pi(b)$ and $\pi(s)$ defined 
in eqs.~\ref{eq:bkgprior} and~\ref{eq:signalprior} are 
improper priors
\footnote{It's easy to show that 
$\int_{0}^{\infty} \pi(b)db = \infty$ and 
$\int_{0}^{\infty} \pi(s) ds = \infty$
for any value of $\alpha$.}
and then they can lead to posterior p.d.f.s that are not
normalizable. In particular, as it will be shown in
sec.~\ref{sec:signalpdf}, this happens when setting
$\alpha \geq 1$. Our calculations will therefore
be valid only for $\alpha<1$. 

The uniform prior for the signal is obtained by 
setting $\alpha=0$ and represents the 
natural choice when the experimenter does not have any
model for the signal. On the other hand, the choice 
of a power-law prior with $\alpha > 0$ will reflect 
the experimenter's belief that small signal values are 
more likely than larger ones.
It is also possible to choose negative values of $\alpha$: 
this choice, that is rather uncommon, would favour 
larger signal values with respect to smaller ones
\footnote{Let $s_{1}<s_{2}$ be two possible signal values
and let us consider the intervals $[s_{1}, s_{1}+\Delta s]$ and
$[s_{2}, s_{2}+\Delta s]$. The ratio between the probabilities 
of finding $s$ in the two intervals is given by
$R = P(s_{1}<s<s_{1}+\Delta s)/P(s_{2}<s<s_{2}+\Delta s)=(s_{2}/s_{1})^{\alpha}$.
When $\alpha>0$ ($\alpha<0$) then $R>1$ ($R<1$) and small (large) 
signal values are more likely than larger (smaller) values. 
For a discussion about the choice of priors in the Bayesian 
approach see for instance the textbook~\cite{cowan}.}.

\subsection{Evaluation of the background posterior p.d.f.}
\label{sec:bkgpdf}

Applying the Bayes'theorem and using for $\pi(b)$ the 
expression in eq.~\ref{eq:bkgprior} it is possible to obtain 
the following equation:

\begin{equation}
p(b | m) = 
\cfrac{p(m|b) \pi(b)}{\int p(m|b) \pi(b) db} =
\cfrac{p(m|b)}{\int_{0}^{\infty}p(m|b)db}
\end{equation}

Finally, replacing $p(m|b)$ with its expression given 
in eq.~\ref{eq:bkgpdf}, it is straightforward to obtain
the final result:

\begin{equation}
\label{eq:bkgposterior}
p(b | m) = \textrm{e}^{-b}~\cfrac{b^{m}}{m!}
\end{equation}

Note that even though the expression of $p(b | m)$ in 
eq.~\ref{eq:bkgposterior} is the same as that of $p(m | b)$ in
eq.~\ref{eq:bkgpdf}, their meanings are completely different. 
In fact, the random variable in eq.~\ref{eq:bkgpdf} is $m$, 
and the formula tells that $m$ follows a Poisson distribution; 
on the other hand, the random variable in 
eq.~\ref{eq:bkgposterior} is $b$, and the formula 
tells that $b$ follows a Gamma distribution.

Using eq.~\ref{eq:bkgposterior} one can also easily evaluate
the average value of $b$ and its standard deviation. It is easy
to show that $\langle b \rangle = m+1$ and 
$\sigma_{b}= \sqrt{m+1}$.

\subsection{Evaluation of the signal posterior p.d.f.}
\label{sec:signalpdf}

The Bayes'theorem can be applied to get the joint posterior p.d.f.
for both $s$ and $b$:

\begin{equation}
\label{eq:jointposterior}
p(s,b|n,m) = 
\cfrac{p(n,m|s,b) \pi(s) \pi(b)}{\int ds \int db p(n,m|s,b) \pi(s) \pi(b)}
\end{equation}

Replacing $p(n,m|s,b)$ with its expression given in 
eq.~\ref{eq:jointpdf} and $\pi(b)$ and $\pi(s)$ with
their expressions given in eqs.~\ref{eq:bkgpdf} 
and~\ref{eq:signalpdf} respectively,
eq.~\ref{eq:jointposterior} can be rewritten as follows:

\begin{equation}
\label{eq:jointposterior2}
p(s,b|n,m) = 
\cfrac{\textrm{e}^{-s-(c+1)b} b^{m} s^{-\alpha} (s+cb)^{n}}
{\int_{0}^{\infty} ds \int_{0}^{\infty} db ~
\textrm{e}^{-s-(c+1)b} b^{m} s^{-\alpha} (s+cb)^{n} }
\end{equation}

Indicating with $N$ the denominator in the right-hand side of 
eq.~\ref{eq:jointposterior2}, it can be rewritten as:

\begin{equation}
\label{eq:N}
N = \int_{0}^{\infty} db ~ b^{m} \textrm{e}^{-(c+1)b}
\int_{0}^{\infty} ds ~\textrm{e}^{-s} s^{-\alpha} (s+cb)^{n} =
\int_{0}^{\infty} db ~ b^{m} \textrm{e}^{-(c+1)b} f(b)
\end{equation}
where we have indicated with $f(b)$ the result of the integral
in $ds$, that can be seen as a function of the variable $b$. 

Applying the binomial theorem, the term $(s+cb)^{n}$ in the 
expression of $f(b)$ can be expanded as follows:

\begin{equation}
(s+cb)^{n} = \sum_{k=0}^{n} \cfrac{n!}{k!(n-k)!} ~s^{k} ~(cb)^{n-k}
\end{equation}

Using this result, the expression of $f(b)$ becomes:

\begin{equation}
f(b) = n! \sum_{k=0}^{n} \cfrac{(cb)^{n-k}}{k!(n-k)!} 
\int_{0}^{\infty} \textrm{e}^{-s} s^{k-\alpha} ds
\end{equation}
and, taking into account the definition of the Gamma function
\footnote{The Gamma function is defined as
$\Gamma(z)=\int_{0}^{\infty} dt \textrm{e}^{-t} t^{z-1}$.
It can be shown that $\Gamma(z)=(z-1)!$ when $z$ is a positive integer.
} the previous equation can be written as:

\begin{equation}
\label{eq:fb}
f(b) = \Gamma(n+1) \sum_{k=0}^{n} (cb)^{n-k} 
\cfrac{\Gamma(k-\alpha+1)}{\Gamma(k+1) \Gamma(n-k+1)}
\end{equation}

Introducing the expression of $f(b)$ given by 
eq.~\ref{eq:fb} in the expression
of $N$ given by eq.~\ref{eq:N} we get the following result:

\begin{equation}
\label{eq:N2}
N = \Gamma(n+1) \sum_{k=0}^{n} 
\cfrac{\Gamma(k-\alpha+1) c^{n-k}}{\Gamma(k+1) \Gamma(n-k+1)}
\int_{0}^{\infty} db~b^{m+n-k} \textrm{e}^{-(c+1)b}
\end{equation}

By making a proper change of variable, the integral in the 
right-hand side eq.~\ref{eq:N2} can be expressed in terms of
a Gamma function. Hence eq.~\ref{eq:N2} can be rewritten as
follows:

\begin{equation}
\label{eq:N3}
N = \cfrac{\Gamma(n+1)}{(c+1)^{m+1}} \sum_{k=0}^{n} 
\cfrac{\Gamma(k-\alpha+1) \Gamma(m+n-k+1)}{\Gamma(k+1) \Gamma(n-k+1)}
\left( \cfrac{c}{c+1} \right)^{n-k}
\end{equation}

The joint posterior p.d.f. for $s$ and $b$ is then given by:

\begin{equation}
\label{eq:jointposterior3}
p(s,b|n,m) = 
\cfrac{1}{N} \textrm{e}^{-s-(c+1)b} b^{m} s^{-\alpha} (s+cb)^{n}
\end{equation}
with the expression of $N$ given in eq.~\ref{eq:N3}.

To evaluate the marginal p.d.f. for $s$ we need to integrate the
joint p.d.f. with respect to $b$:

\begin{equation}
\label{eq:signalposterior}
p(s|n,m) = \int_{0}^{\infty} p(s,b|n,m)db = 
\cfrac{1}{N}~\textrm{e}^{-s} s^{-\alpha} 
\int_{0}^{\infty} db~\textrm{e}^{-(c+1)b}b^{m}(s+cb)^{n}
\end{equation} 

Indicating with $g(s)$ the integral in the right-hand side of
eq.~\ref{eq:signalposterior}, it can be evaluated in a similar
way to that used to calculate $f(b)$. It is easy to show that:

\begin{equation}
\label{eq:gs}
g(s) = \cfrac{\Gamma(n+1)}{(c+1)^{m+1}} \sum_{k=0}^{n} 
\cfrac{\Gamma(m+n-k+1)}{\Gamma(k+1) \Gamma(n-k+1)}
\left( \cfrac{c}{c+1} \right)^{n-k}~s^{k}
\end{equation}

Introducing in eq.~\ref{eq:signalposterior} the expression of 
$g(s)$ given by eq.~\ref{eq:gs} and the expression of $N$ given 
by eq.~\ref{eq:N3}, it is possible to show that the posterior 
p.d.f. for $s$ is given by:

\begin{equation}
\label{eq:signalposterior2}
p(s|n,m) = \cfrac{
\sum\limits_{k=0}^{n} \cfrac{\Gamma(m+n-k+1)}{\Gamma(k+1) \Gamma(n-k+1)} 
\left( \cfrac{c}{c+1} \right)^{n-k} s^{k-\alpha} \textrm{e}^{-s}
}{
\sum\limits_{k=0}^{n} \cfrac{\Gamma(k-\alpha+1) \Gamma(m+n-k+1)}{\Gamma(k+1) \Gamma(n-k+1)} 
\left( \cfrac{c}{c+1} \right)^{n-k} 
}
\end{equation}

It is worth to point out here that eq.~\ref{eq:signalposterior2} 
is valid only for $\alpha < 1$. In fact, if $\alpha \geq 1$, 
the argument $k-\alpha+1$ of the Gamma function in the sum in 
the right-hand side of the denominator will be null or negative, 
and consequently the Gamma function either will assume
negative values or will diverge. As discussed in sec.~\ref{sec:priors}, 
this behaviour is due to the fact that since the posterior 
p.d.f. $p(s|n,m)$ is obtained starting from improper 
priors, its normalization is not guaranteed.

\begin{figure}[ht]
\begin{center}
\includegraphics[width=1.00\textwidth]{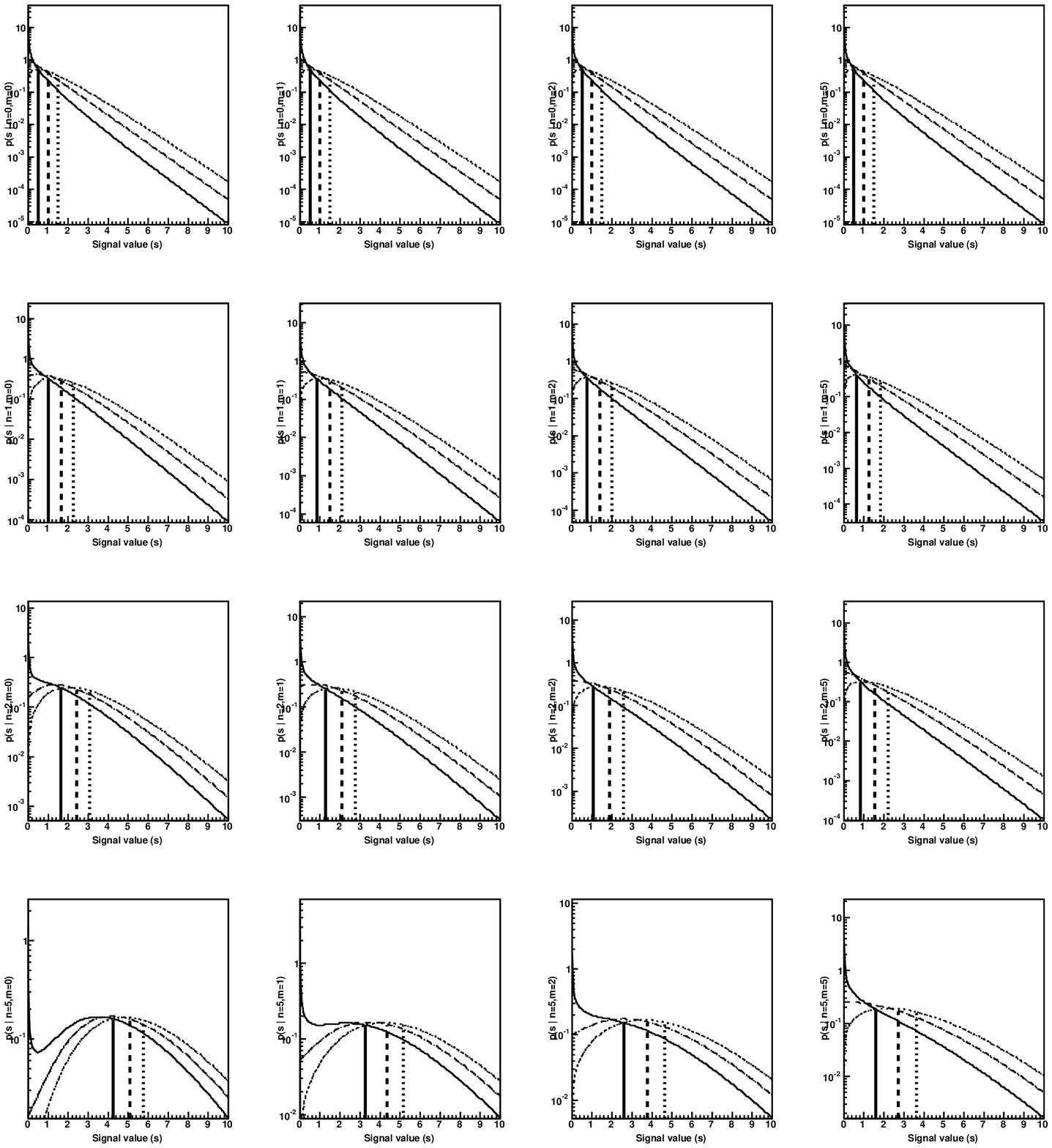}
\end{center}
\caption{The plots show the posterior p.d.f.s for the 
signal $p(s|n,m)$ for different values of $n$,$m$ and $\alpha$
in the case $c=1$. Going from top to bottom, the plots 
correspond to $n=0,1,2,5$; going from left to right the plots
correspond to $m=0,1,2,5$. The p.d.f.s have been evaluated
assuming $\alpha=0.5$ (continuous lines), $\alpha=0$ (dashed lines)
and $\alpha=-0.5$ (dotted lines). In each plot three vertical lines
are also drawn with the same styles specified above, each 
indicating the average value of $s$ from the corresponding p.d.f.s.}
\label{fig:pdfplots}
\end{figure}

Fig.~\ref{fig:pdfplots} shows the posterior p.d.f.s for the
signal $s$ evaluated for some different values of $n$ and $m$
in the case $c=1$, i.e. when the background region has the 
same size as the signal region. The calculations have been
performed in the cases $\alpha=0.5$ (small signal expected), 
$\alpha=0$ (uniform prior) and $\alpha=-0.5$ (large signal
expected). The differences between the three p.d.f.s become
negligible when when $n$ is larger than $cm$. 

From eq.~\ref{eq:signalposterior2} one can also easily 
calculate the moments of the p.d.f. $p(s|n,m)$. It's easy
to show that the $h$th moment of the p.d.f. is given by:

\begin{equation}
\label{eq:moments}
\langle s^{h} | n,m \rangle =  \cfrac{
\sum\limits_{k=0}^{n} 
\cfrac{\Gamma(k-\alpha+1+h) \Gamma(m+n-k+1)}{\Gamma(k+1) \Gamma(n-k+1)} 
\left( \cfrac{c}{c+1} \right)^{n-k} 
}{
\sum\limits_{k=0}^{n} 
\cfrac{\Gamma(k-\alpha+1) \Gamma(m+n-k+1)}{\Gamma(k+1) \Gamma(n-k+1)} 
\left( \cfrac{c}{c+1} \right)^{n-k} 
}
\end{equation}

In particular, eq.~\ref{eq:moments} allows to calculate
the expectation value of $s$, i.e. $\langle s \rangle$ and
its variance, that can be evaluated as 
$\textrm{var}(s)=\langle s^{2} \rangle - \langle s \rangle^{2}$.  

Finally, we can consider the limiting case with absence 
of background, that can be obtained by setting $c=0$
(background region with null size). In this limit, 
all the terms with $[c/(c+1)]^{n-k}$ in both the summations 
of eq.~\ref{eq:signalposterior2} will vanish, but the ones 
with $k=n$, where $[c/(c+1)]^{n-k} = 1$. Hence, in this
case the posterior p.d.f. for the signal simplifies to:

\begin{equation}
\label{eq:signalposterior3}
p(s|n,m) = \cfrac{s^{n-\alpha} \textrm{e}^{-s}}{\Gamma(n-\alpha+1)}
\end{equation}

and, as expected, does not depend on $m$.
In particular, in the case $\alpha=0$ the posterior 
p.d.f. for the signal becomes a Gamma distribution. 

\section{Evaluation of the upper limits on the signal}
\label{sec:upperlimits}

The posterior p.d.f. for the signal given by 
eq.~\ref{eq:signalposterior2} can be used to evaluate 
Bayesian confidence intervals for $s$.
In particular, in this section, we will apply the 
result of eq.~\ref{eq:signalposterior2} to the calculation
of upper limits on $s$. 

To evaluate the upper limit $s_{u}$ at the confidence level
$1-\beta$ we have to solve the integral equation: 

\begin{equation}
\label{eq:uldef}
\int_{0}^{s_{u}} p(s|n,m)ds = 1 - \beta
\end{equation}

Taking advantage of the fact that

\begin{equation}
\int_{0}^{s_{u}} s^{k-\alpha} \textrm{e}^{-s} ds = 
\gamma(k-\alpha+1, s_{u})
\end{equation}
where we have indicated with $\gamma$ the incomplete Gamma function
\footnote{The incomplete Gamma function is defined as 
$\gamma(z,x)=\int_{0}^{x}dt \textrm{e}^{-t} t^{z-1}$. According to
this definition $\gamma(z,\infty)=\Gamma(z)$.},
eq.~\ref{eq:uldef} can be rewritten as:

\begin{equation}
\label{eq:upperlimit}
1 - \beta = \cfrac{
\sum\limits_{k=0}^{n} 
\cfrac{\gamma(k-\alpha+1,s_{u}) \Gamma(m+n-k+1) }{\Gamma(k+1) \Gamma(n-k+1)} 
\left( \cfrac{c}{c+1} \right)^{n-k} 
}{
\sum\limits_{k=0}^{n} 
\cfrac{\Gamma(k-\alpha+1) \Gamma(m+n-k+1)}{\Gamma(k+1) \Gamma(n-k+1)} 
\left( \cfrac{c}{c+1} \right)^{n-k} 
}
\end{equation}

Eq.~\ref{eq:upperlimit} can be solved numerically and allows
to obtain the Bayesian upper limits on $s$ at the confidence level
$1-\beta$ for any values of $n$ and $m$, given the values of
$c$ and $\alpha$. We have performed our calculations
using the CERN ROOT package~\cite{ROOT}. In particular,
we have implemented a code that allows to evaluate 
the numerical solutions $s_{u}$ of eq.~\ref{eq:upperlimit} 
for any given value of $\beta$ with 
the Brent's method, using the ROOT built-in tools. 

\begin{figure}[ht]
\begin{center}
\includegraphics[width=0.6\textwidth]{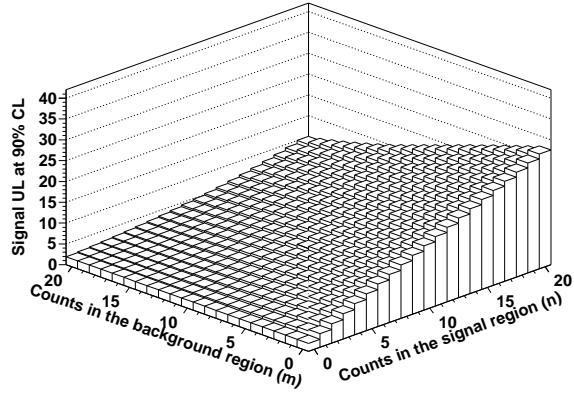}
\includegraphics[width=0.6\textwidth]{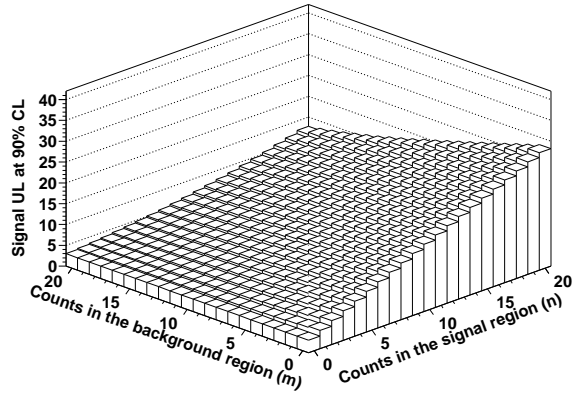}
\includegraphics[width=0.6\textwidth]{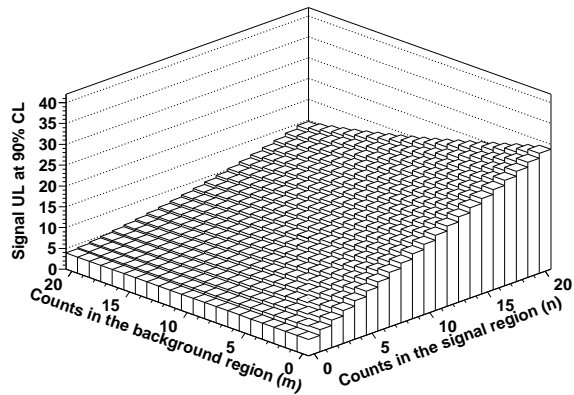}
\end{center}
\caption{The plots show the values of the signal upper limits at $90\%$
confidence level as a function of the counts observed in the signal 
and background regions, $n$ and $m$, in the case $c=1$. Going from
top to bottom, the plots correspond to $\alpha=0.5$, $\alpha=0$ and
$\alpha=-0.5$.}
\label{fig:ulsummary}
\end{figure}

In fig.~\ref{fig:ulsummary} the upper limits on the signal 
at $90\%$ confidence level are shown as a function of the
observed counts in the signal ($n$) and background ($m$) 
regions in the case $c=1$ for three different values of $\alpha$. 
As expected, the choice of the power law index $\alpha$ in the signal 
prior p.d.f. will affect the result on the upper limits. 
In particular, the upper limits on the signal will increase 
with decreasing $\alpha$. This result is in agreement with 
the fact that when $\alpha$ is positive and close to $1$ small 
signal values are expected while, on the other hand, when $\alpha$ 
is negative, large signal values are expected. 
As pointed out in sec.~\ref{sec:intro}, this is a general issue
of the Bayesian approach, the results being influenced by the 
initial belief of the experimenter. 

\begin{figure}[ht]
\begin{center}
\includegraphics[width=1.0\textwidth]{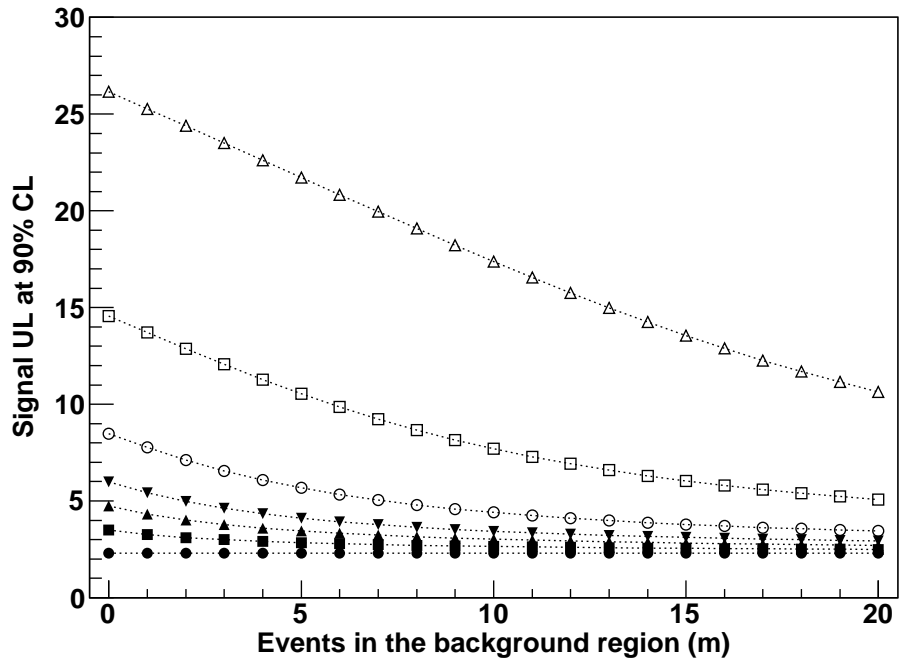}
\end{center}
\caption{The plot shows the values of the signal upper limits at $90\%$
confidence level as a function of the counts observed in the background region, 
$m$, in the case $c=1$ and $\alpha=0$, for some different values of the counts 
in the signal region. The calculation has been performed for 
$n=0$ ($\bullet$), $1$ ($\blacksquare$), $2$ ($\blacktriangle$),
$3$ ($\blacktriangledown$), $5$ ($\circ$), $10$ ($\square$)
and $20$ ($\vartriangle$).}
\label{fig:ulfixedn}
\end{figure}

Fig.~\ref{fig:ulfixedn} shows the upper limits on the
signal at $90\%$ confidence level as a function of the
background events $m$ for several values of the signal
events $n$. The calculation has been performed in
the case $c=1$ with the uniform prior ($\alpha=0$).
For any given value of $n$, the upper limit on 
the signal decreases with increasing $m$, in
agreement with the fact that (in the cases when
$n>cm$) a rough estimate of $s$ is given by 
$n-cm$ and the upper limit is expected to be 
proportional to $n-cm$. It has also to be
pointed out that in the case $n=0$, i.e. when 
no events are observed in the signal region, the 
upper limit is always equal to $2.30$, independently
on $m$
\footnote{It is evident that when no events 
(or a few events) are observed in the signal region 
and a large number of events are observed in the 
background region the prior p.d.f. given by 
eq.~\ref{eq:signalprior} is not adequate. 
In these cases a different prior p.d.f. should be used,
that allows for negative values of $s$.}. 

\begin{figure}[ht]
\begin{center}
\includegraphics[width=1.0\textwidth]{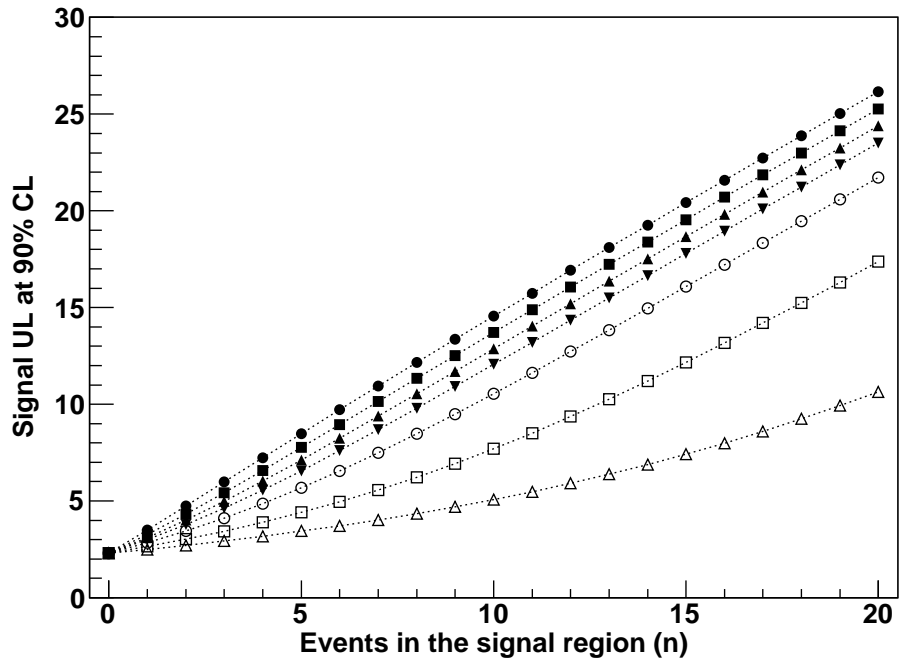}
\end{center}
\caption{The plot shows the values of the signal upper limits at $90\%$
confidence level as a function of the counts observed in the signal region, 
$n$, in the case $c=1$ and $\alpha=0$, for some different values of the counts 
in the background region. The calculation has been performed for 
$m=0$ ($\bullet$), $1$ ($\blacksquare$), $2$ ($\blacktriangle$),
$3$ ($\blacktriangledown$), $5$ ($\circ$), $10$ ($\square$)
and $20$ ($\vartriangle$).}
\label{fig:ulfixedm}
\end{figure}

Finally, fig.~\ref{fig:ulfixedm} shows the upper limits on the
signal at $90\%$ confidence level as a function of the
signal events $m$ for several values of the background
events $m$. Again, the calculation has been performed in
the case $c=1$ with the uniform prior ($\alpha=0$).
The upper limits increase with increasing
$n$, and the trend of the curves becomes almost
linear when $n>cm$. 

\subsection{Study of the frequentist coverage}

To study the frequentist coverage of the upper 
limits obtained with our procedure we implemented 
a dedicated simulation. For simplicity we studied 
only the case with $c=1$, when the signal and the 
background regions have the same sizes.

For any given pair of values of $s$ and $b$, a sample 
of $10^{5}$ events was simulated, each corresponding to 
the outcome of an experiment. Each event consists of 
a couple of random integer numbers $(n,m)$, representing 
respectively the counts in the signal and in the background 
regions, that are generated according to the p.d.f. 
of eq.~\ref{eq:jointpdf}.
For each couple of counts $(n,m)$ the corresponding 
upper limits on $s$ at $90\%$ confidence level were 
evaluated by solving eq.~\ref{eq:upperlimit} for
different values of $\alpha$. The coverage was then
evaluated as the fraction of events with an associated 
upper limits less than the true value of the signal $s$. 

\begin{figure}[ht]
\begin{center}
\includegraphics[width=0.55\textwidth]{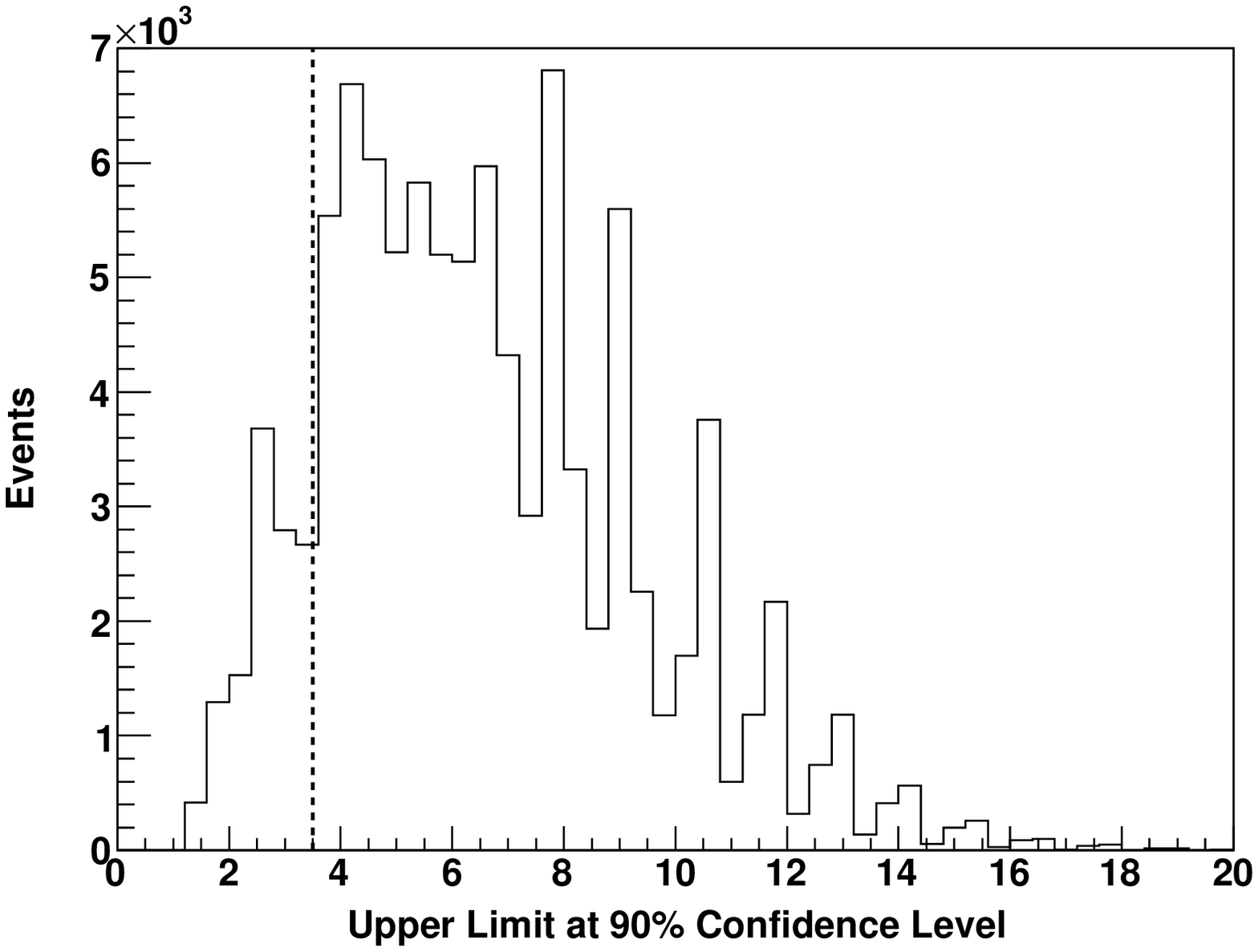}
\includegraphics[width=0.55\textwidth]{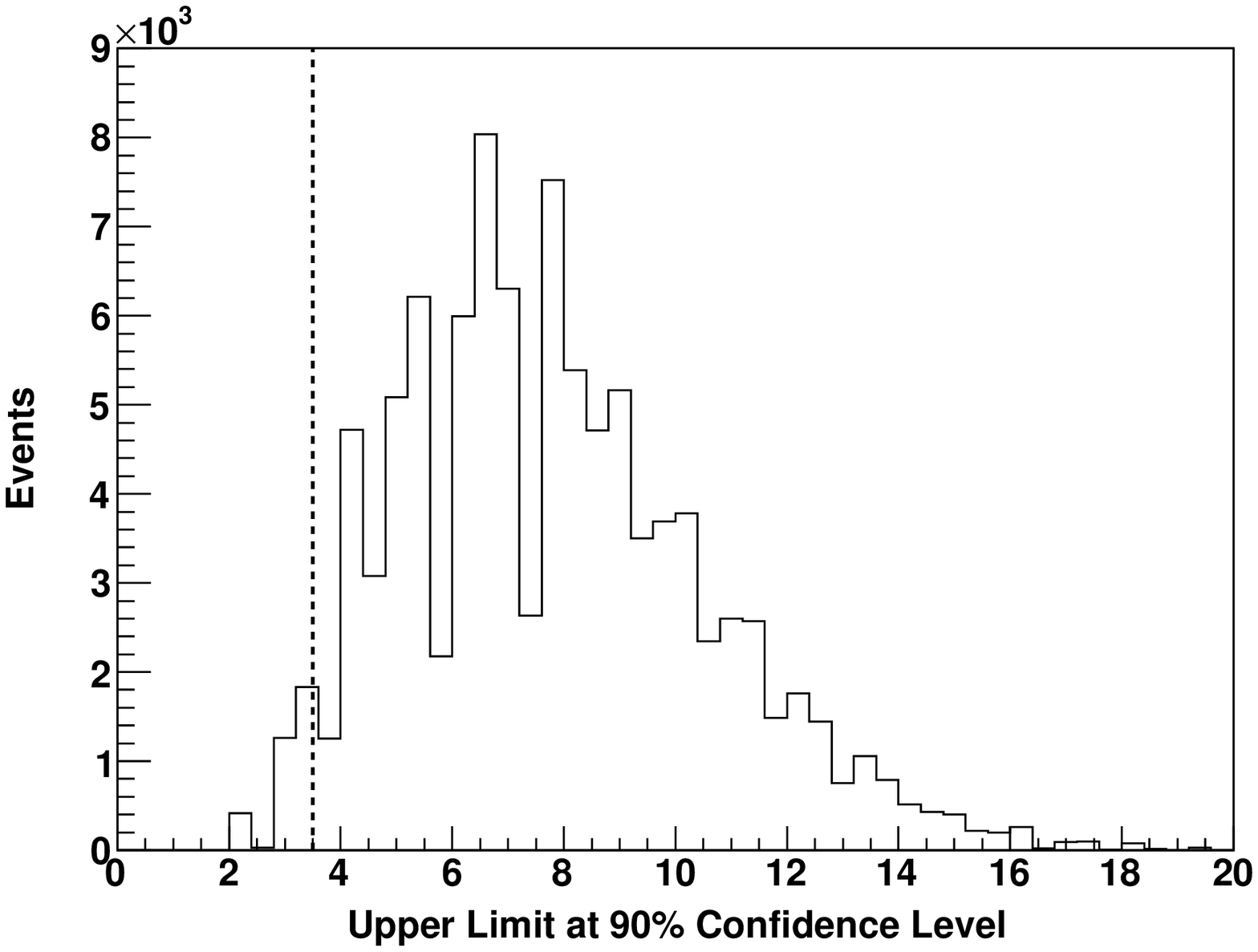}
\includegraphics[width=0.55\textwidth]{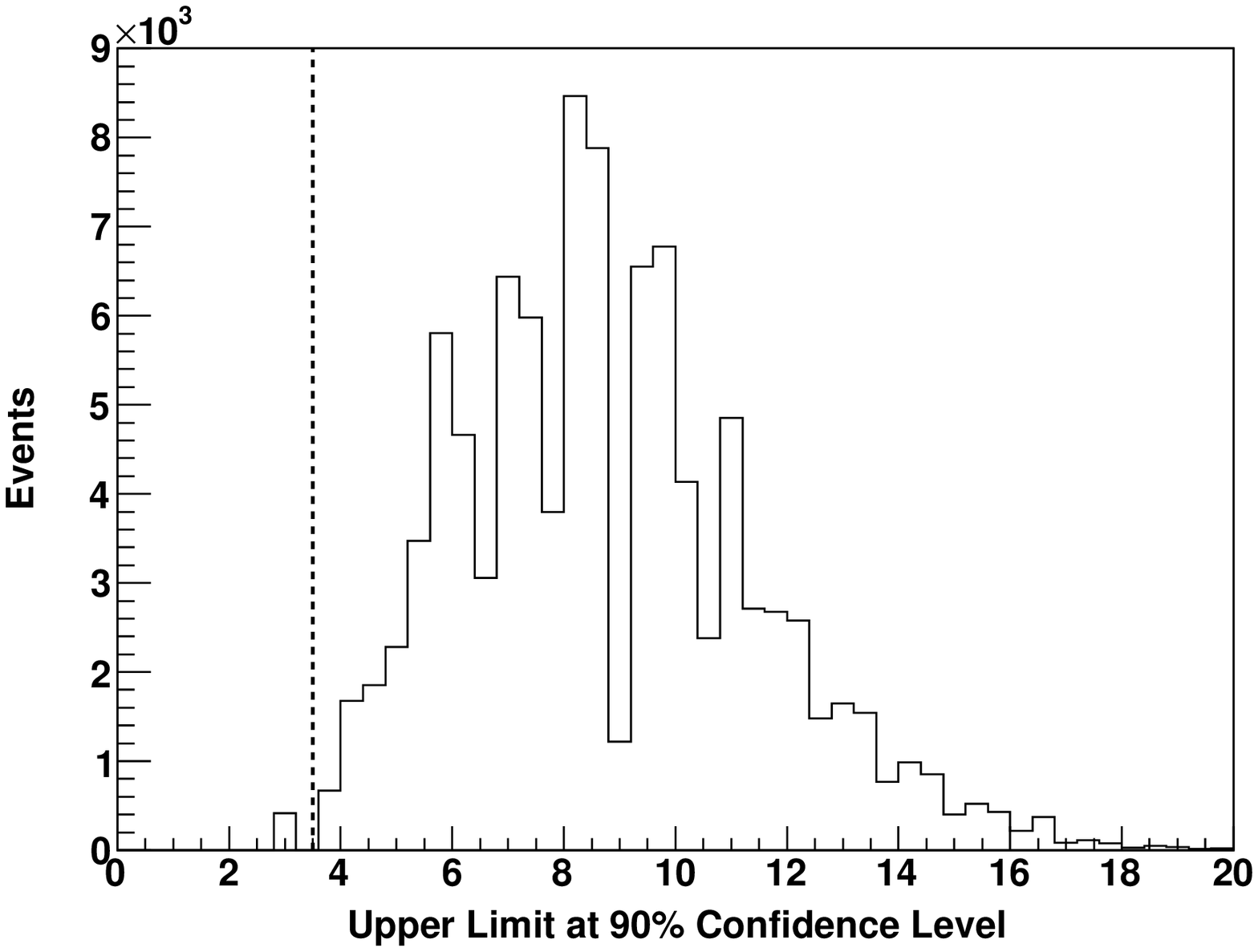}
\end{center}
\caption{The plots show the distributions of the signal upper 
limits at $90\%$ confidence level obtained from a sample of 
$10^{5}$ simulated experiments with $s=3.5$, $b=2$ and $c=1$.
Going from top to bottom, the plots correspond to the upper
limits evaluated by setting $\alpha=0.5$, $\alpha=0$ and
$\alpha=-0.5$. The dashed lines indicate the true value of the
signal. The coverage is graphically represented by the area 
of the histogram at the right of the dashed line normalized 
to the total area.}
\label{fig:coverageexample}
\end{figure}

Fig.~\ref{fig:coverageexample} shows, as an example, 
the results obtained with a simulated sample of events 
with $s=3.5$ and $b=2$. The distributions of the upper 
limits at $90\%$ confidence level are shown for 
$\alpha=0.5$, $\alpha=0$ and $\alpha=-0.5$. As expected, the 
coverage increases with decreasing $\alpha$, since
lower values of $\alpha$ correspond to more conservative 
upper limits. 

\begin{figure}[ht]
\begin{center}
\includegraphics[width=0.45\textwidth]{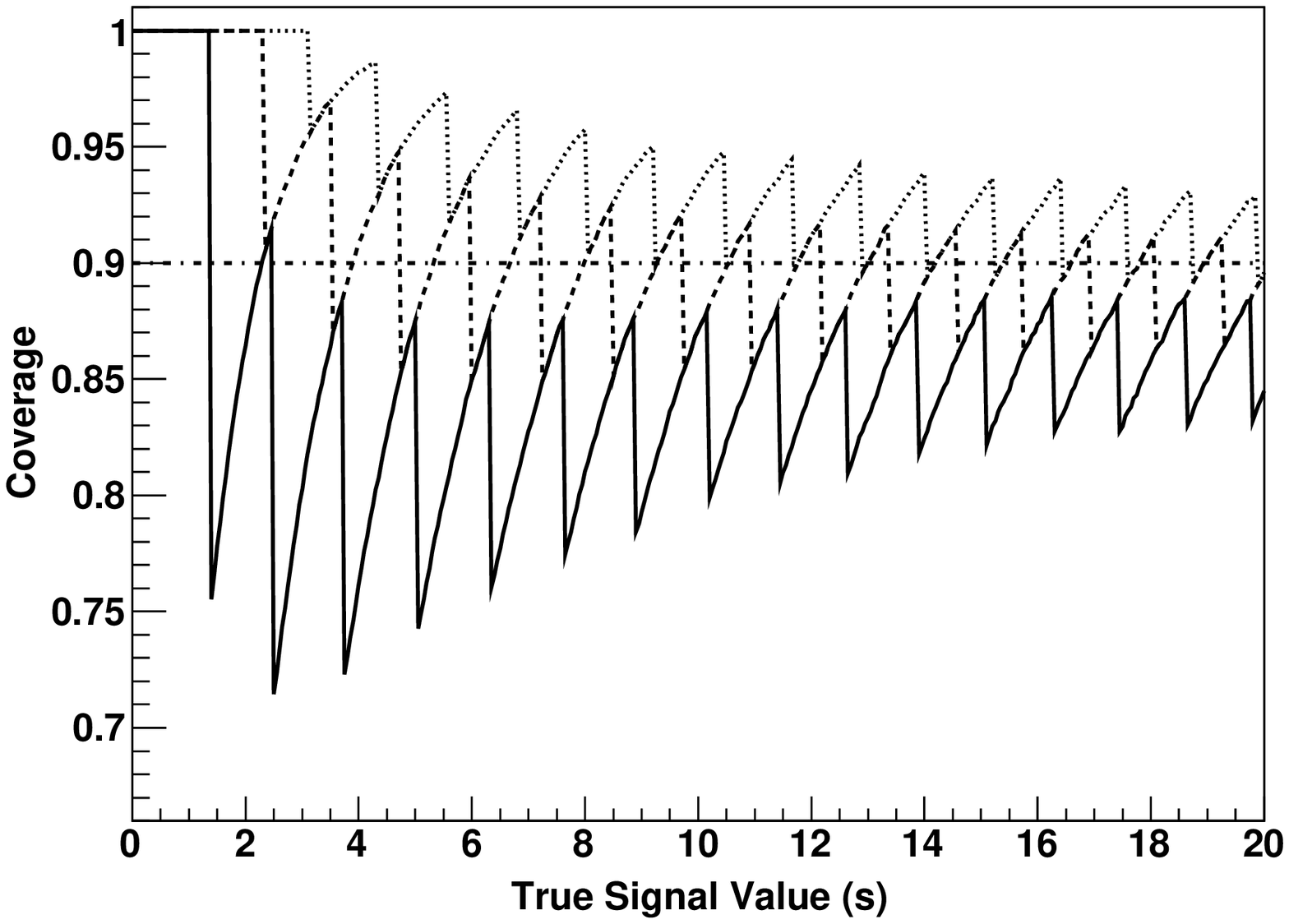}
\includegraphics[width=0.45\textwidth]{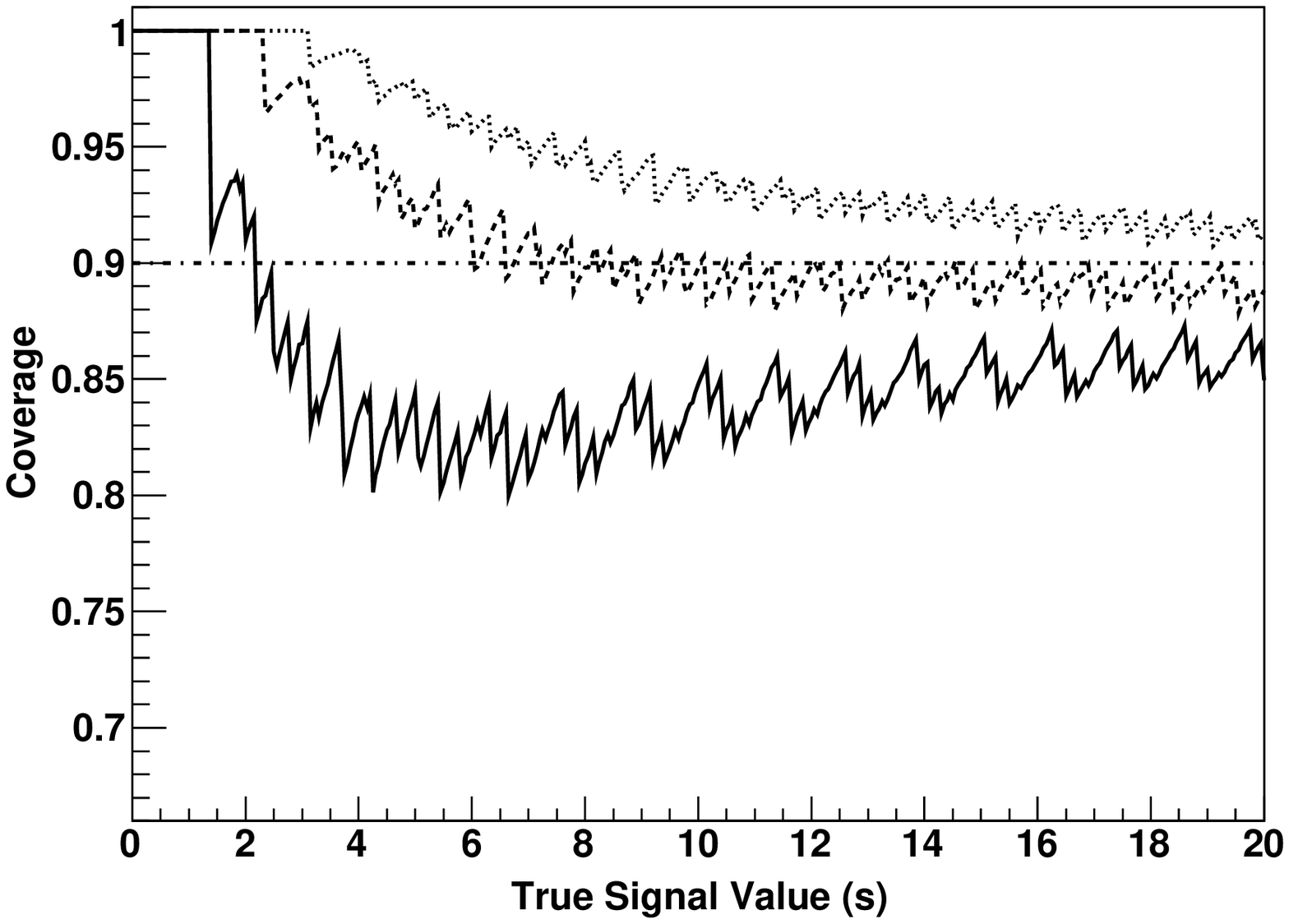}
\includegraphics[width=0.45\textwidth]{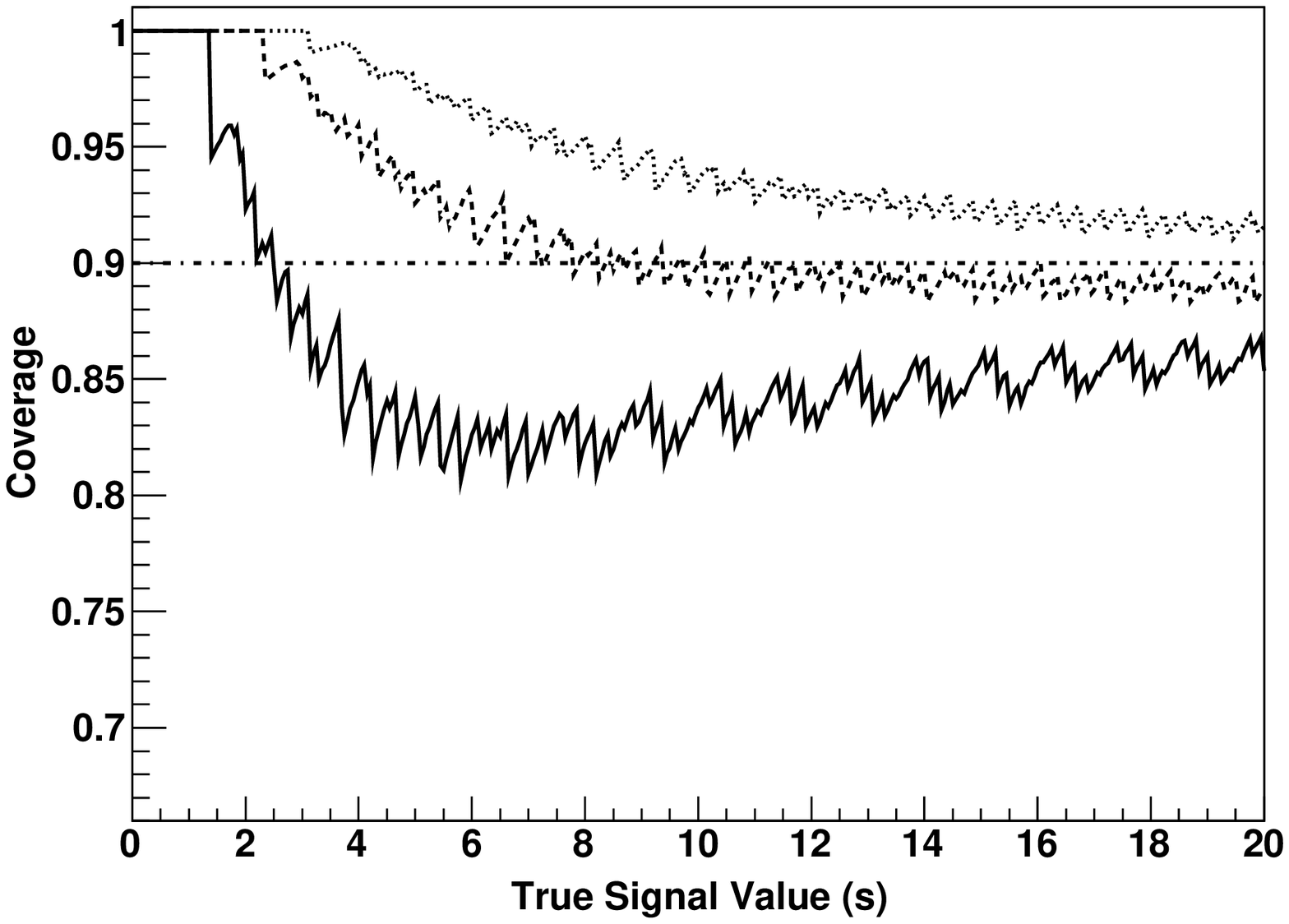}
\includegraphics[width=0.45\textwidth]{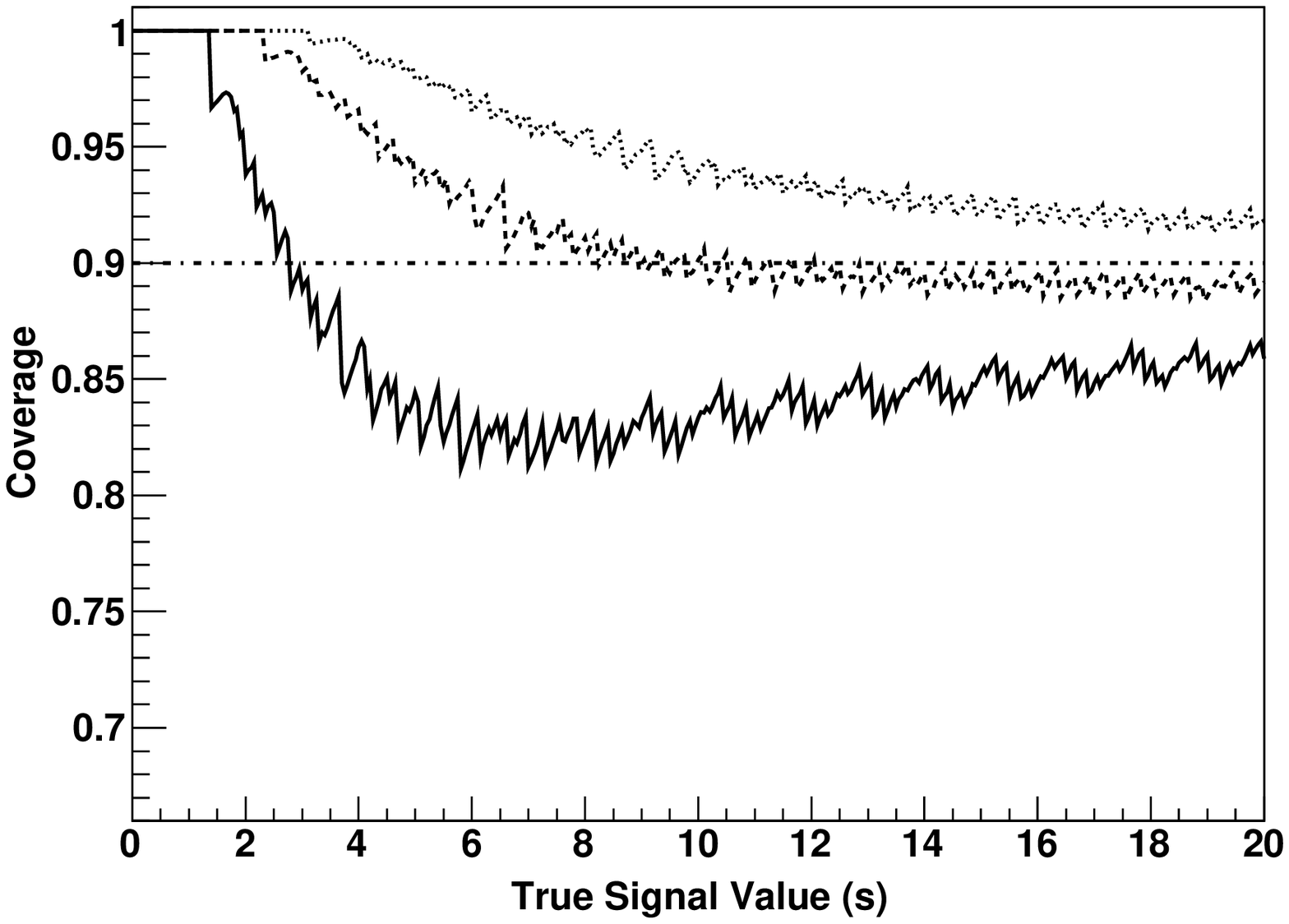}
\includegraphics[width=0.45\textwidth]{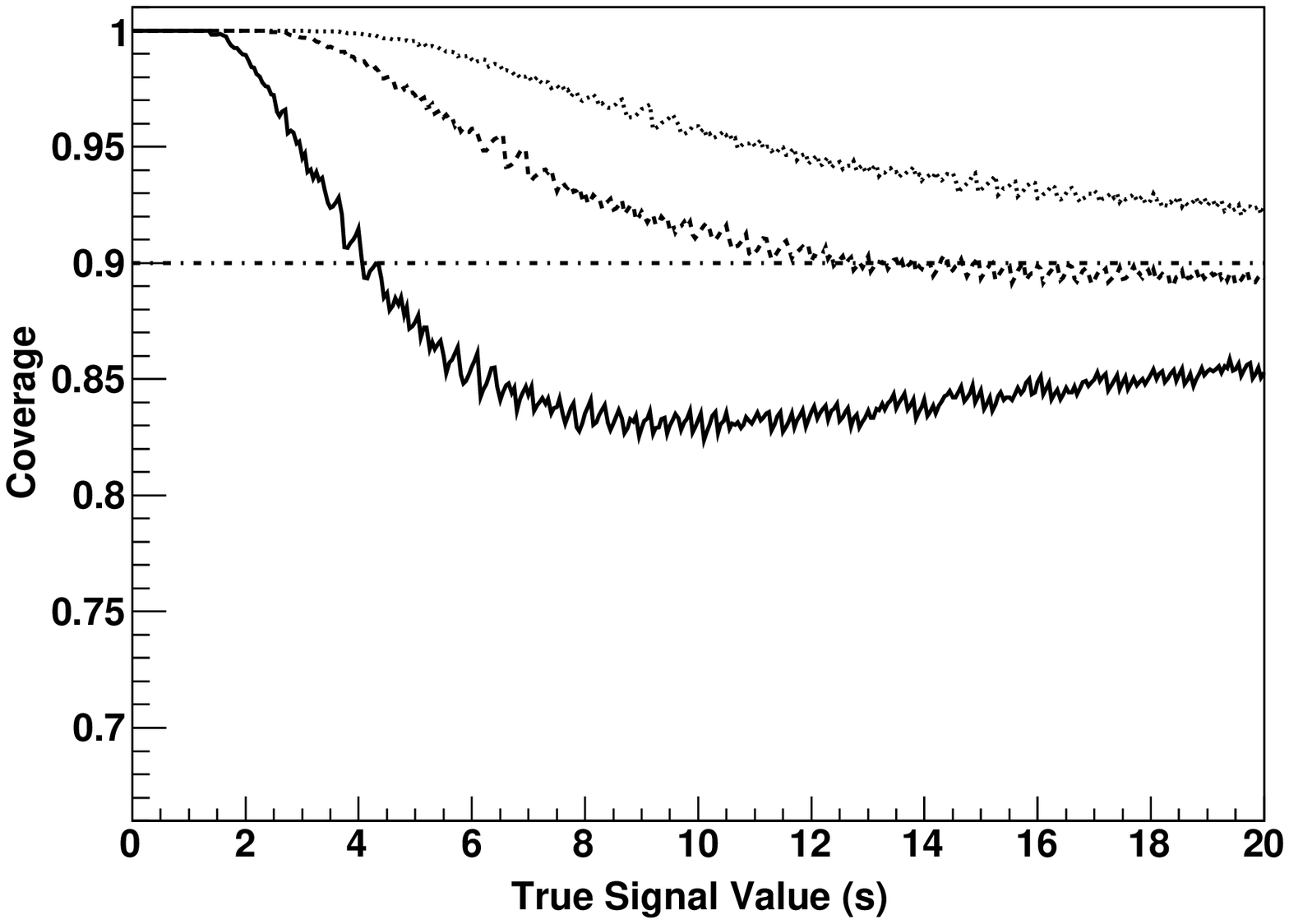}
\includegraphics[width=0.45\textwidth]{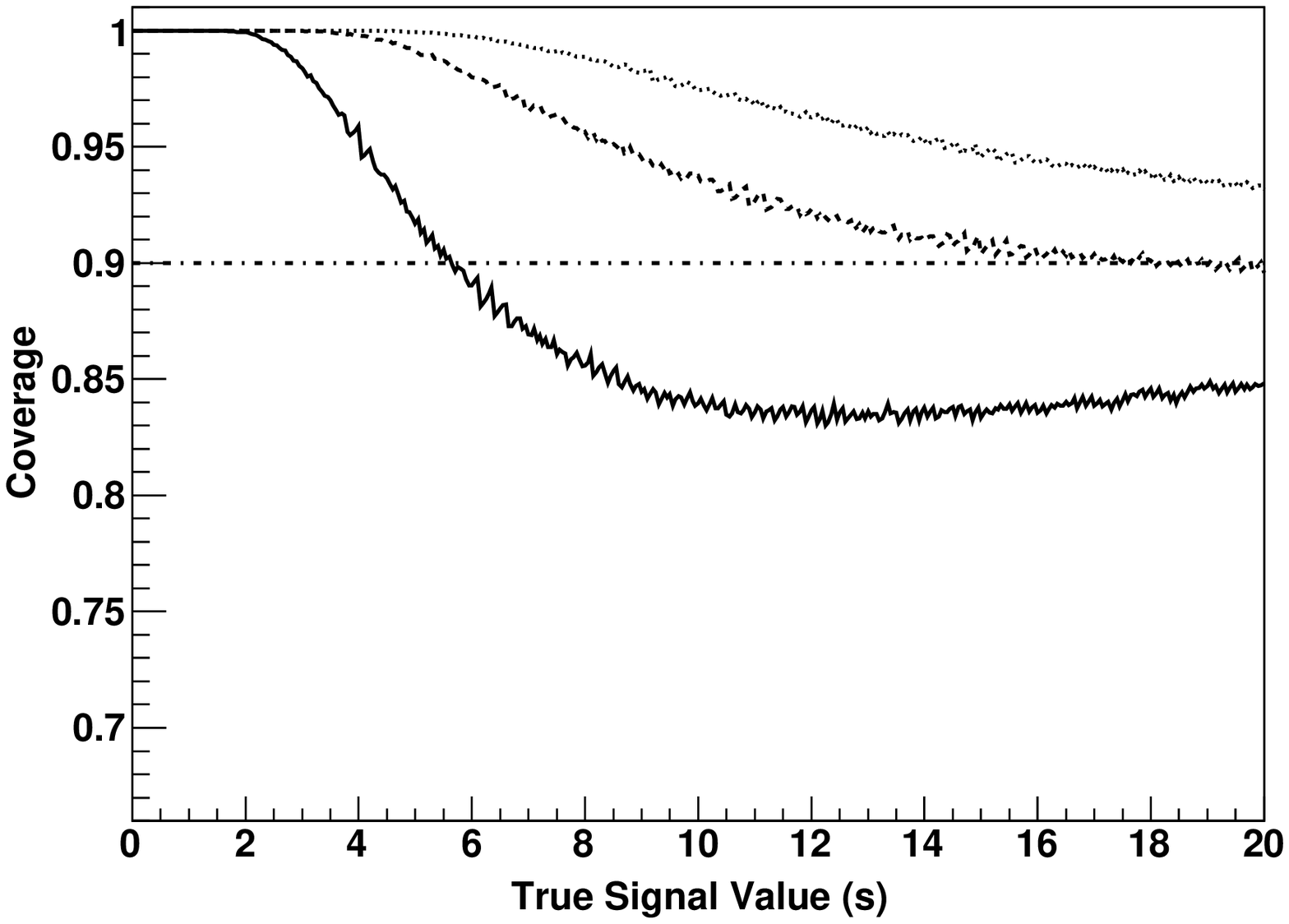}
\end{center}
\caption{The plots show the frequentist coverage 
of the upper limits at $90\%$ confidence level 
as a function of the true signal value $s$ for
different values of $b$ with $c=1$. 
From top left to bottom right the plots correspond 
to the cases $b=0$, $b=1$, $b=1.5$, $b=2$, $b=5$ and $b=10$.
The upper limits have been evaluated by setting 
$\alpha=0.5$ (continuous lines), $\alpha=0$ (dashed lines) 
and $\alpha=-0.5$ (dotted lines). The dash-dotted lines 
indicate the $90\%$ coverage.}
\label{fig:coveragesummary}
\end{figure}

We have studied the frequentist coverage of our upper
limits as a function of both $s$ and $b$.
The results are summarized in  Fig.~\ref{fig:coveragesummary}, 
where the coverage is plotted as a function of $s$ in 
the cases when $b=0$,$b=1$,$b=1.5$,$b=2$,$b=5$ and $b=10$.  
As expected, the frequentist coverage decreases
with increasing $\alpha$. The choice of a uniform prior
for the signal, i.e. $\alpha=0$ guarantees a coverage
that is larger than $90\%$ for low signal values,
and oscillates around $90\%$ with increasing $s$.
On the other hand, the choice of a less conservative 
prior with $\alpha=0.5$ does not seems to affect significantly 
the coverage in case of an high background level.

\section{Conclusions}

We have developed a procedure that, following the
Bayesian approach, allows to evaluate confidence intervals
on the signal in experiments with background where 
both signal and background are modeled by the Poisson
statistics. The implementation of the  method is 
quite simple from the mathematical point of view, 
and does not require any adjustments to treat the 
cases when less events are observed in the 
signal region than those in the background region.
The results obtained when our procedure is applied
to the calculation of upper limits have been also
discussed. The frequentist coverage of the 
upper limits evaluated with this procedure has 
been also studied.


\begin{thebibliography}{99}

\bibitem{pdg} G.~Cowan et al. (Particle Data Group), {\em Journ.~Phys. G} 
{\bf 37} (2010), 075021 see also
http://pdg.lbl.gov/2010/reviews/rpp2010-rev-statistics.pdf

\bibitem{neyman} J.~Neyman, {\em Phil. Trans. Royal Soc. London, Series A}
{\bf 236}, 333 (1937)

\bibitem{feldman} G.~J.~Feldman, R.~D.~Cousins, ``Unified approach
to the classical statistical analysis of small signals'' ,
{\em Phys. Rev. D} {\bf 57} (1998), 3873

\bibitem{rolke1} W.~A.~Rolke, A.~M.~Lopez,
``Confidence intervals and upper bounds for small signals in
presence of background noise'',
{\em Nucl. Inst. Meth.} {\bf A458} (2000), 745

\bibitem{rolke2} W.~A.~Rolke, A.~M.~Lopez, J.~Conrad, 
``Limits on Confidence Intervals in the Presence of
Nuisance Parameters'', {\em ArXiv:physics/0403059} (2009)

\bibitem{cowan} G.~Cowan, ``Statistical Data Analysis'', Oxford University Press (1998), 
ISBN 0-19-850155-2

\bibitem{helene} O.~Helene, 
``Determination of the upper limit of a peak area'',
{\em Nucl. Inst. Meth.} {\bf 212} (1983), 319

\bibitem{heinrich} J.~Heinrich et al., ``Interval estimation in the
presence of nuisance parameters. Bayesian approach'',
{\em ArXiv:physics/0409129} (2004)

\bibitem{ROOT} R.~Brun anf F.~Rademakers,
``ROOT - An object oriented data analysis framework'',
{\em Nucl. Inst. Meth.} {\bf A389} (1997), 81 (see also
http://root.cern.ch)

\end{thebibliography}
\end{document}